# The Star Formation Histories and Efficiencies of Two Giant HII Regions in M33

Christine D. Wilson and Brenda C. Matthews[1]

Department of Physics and Astronomy, McMaster University, Hamilton, Ontario, Canada

L8S 4M1

astro-ph/9506104   20 Jun 95

[1]Also Department of Physics and Astronomy, University of Calgary, Calgary, Alberta, Canada T2N 1N4

## ABSTRACT

We have used $UBV$ photometry to re-identify the OB associations which power the two most luminous HII regions in M33, NGC 604 and NGC 595. Color-magnitude diagrams of the two OB associations reveal a significant difference (2-3 Myr) in the ages of the most recent star formation episode in these two regions. In addition, the presence of evolved, low-mass red supergiants in NGC 595 shows that this region has undergone a prior episode of star formation $\sim 10 - 15$ Myr ago. These data, combined with the presence of molecular clouds in the heart of NGC 604, suggest that molecular clouds may survive at least one intense episode of massive star formation. The different star formation histories of the two regions provide a good explanation for their different gas structure: in NGC 595 the interstellar medium is primarily atomic, since the massive stars have had enough time to photo-dissociate most of the molecular gas, while the younger NGC 604 is more typical of star-forming regions and is dominated by molecular gas. The number of Wolf-Rayet and other evolved stars is consistent with the estimated turn-off ages and the number of stars still on the main sequence. The star formation efficiencies (mass of stars per mass of gas) of these two HII regions are up to a factor of 3 larger than the average efficiency in the inner disk of M33 or in Galactic molecular clouds, but are still only 2-5%. The very large $H\alpha$ luminosities of these regions appear to be a product of the increased star formation efficiency, the large gas reservoir which allows a much larger number of massive stars to be formed, and the tendency for a young co-eval stellar population to produce more ionizing photons than a steady-state population of similar mass.

*Subject headings:* Galaxies: Individual (M33) – Galaxies: Stellar Content –

ISM: HII Regions – ISM: Molecules – ISM: Individual (NGC 604, NGC 595) – Local Group – Stars: Formation

## 1. Introduction

Giant HII regions are the most spectacular examples of massive star formation seen in normal galaxies. Such regions require typically tens to hundreds of O stars to provide enough ionizing radiation to produce a large ionized gas region and are both larger and brighter than normal Galactic HII regions. The structural properties of the ionized gas appear very similar in HII regions covering a wide range of luminosities, which suggests that HII regions (including the so-called "super-giant" HII regions) follow a continuous distribution of physical properties (Kennicutt 1984). However, the ionized gas may have undergone substantial processing by the ionizing star cluster, and may no longer be representative of the initial conditions in the gas in the cluster. Differences in the physical conditions, such as the initial gas mass and the star formation efficiency (mass of stars formed per mass of gas), between normal and giant HII regions are likely to be important for understanding how and why these massive regions form. Measuring the massive star content of giant HII regions is important for determining the star formation efficiency of the region, which may provide a clue to whether giant HII regions are physically distinct from normal, less luminous HII regions, or whether they simply represent the high-luminosity tail of the HII region distribution. However, the stellar content of these interesting regions can only be studied directly in galaxies within the Local Group (i.e. 30 Doradus, Parker & Garmany 1993). In more distant systems, the stellar content of giant HII regions must be deduced using spectral synthesis techniques, for which it is difficult to sort out competing effects of age, metallicity, and changes in the initial mass function (Rosa & Benvenuti 1994).

The most luminous HII regions in the Local Group after 30 Doradus are located in the spiral galaxy M33 (Kennicutt 1984). The two brightest of these, NGC 604 and NGC 595, have H$\alpha$ luminosities that are lower than that of 30 Doradus by factors of 3 and 10, respectively. The gas content and star formation properties of these two regions have been measured using CO, HI, and H$\alpha$ data (Wilson & Scoville 1992, Kennicutt 1988). The two regions were found to contain comparable amounts of gas to similar sized but less active regions in the inner disk of M33, although both regions had a somewhat higher atomic gas mass fraction than normal. In contrast to the inner disk regions, these 100-200 pc size gas complexes appear to be gravitationally bound. The gas depletion timescales, commonly defined to be the mass of gas divided by the star formation rate, obtained from the H$\alpha$ luminosity and the total gas mass are an order of magnitude smaller in the giant HII regions than in the inner disk, which suggests that these regions have an order of magnitude larger star formation efficiency. However, the star formation rates were calculated using a steady-state model, i.e., one in which the star formation rate is assumed constant with time, while these regions are more likely to be characterized by a single burst of star formation. Thus the star formation rates derived in Wilson & Scoville (1992) may be incorrect and may lead to a false picture of the star formation efficiency in these regions.

At most wavelengths, NGC 595 appears to be a scaled-down version of NGC 604, with a luminosity a factor of 3 smaller (Wilson & Scoville 1992). However, there are two intriguing differences between the HII regions. The interstellar medium in NGC 595 is predominately atomic, unlike that in NGC 604 and star-forming regions in the inner disk of M33 where molecular gas is the dominant form. Although the unusually high atomic gas content of both regions suggests that photo-dissociation of molecular hydrogen by ultraviolet photons to produce atomic hydrogen is more important in these HII regions than in less active regions of M33, it appears to have been most effective in NGC 595. Perhaps NGC 595 is older than NGC 604 and hence has been more effective at photo-dissociating its molecular gas. In addition, the number of Wolf-Rayet stars, the evolved descendents of massive stars, compared to main-sequence O stars is a factor of 3 higher in NGC 595 than in NGC 604 (Drissen, Moffat, & Shara 1993). Differences between the two regions that could account for this observation include the age, the form of the initial mass function, and the duration of the burst. Drissen et al. (1993) suggest that NGC 595 has experienced a single burst of star formation, while NGC 604 has undergone more than one episode of star formation.

We can determine the age of the OB associations and look for evidence of prior episodes of star formation using broad-band photometry of the blue stars. Comparing the location of the turn-off with evolutionary tracks or isochrones will give the age of the most recent burst of star formation, while the evolved, lower-mass red supergiants often seen in OB associations (e.g. Massey et al. 1989) can provide evidence for a prior episode of star formation. Massive star formation rates for the regions can also be obtained directly from star counts of the high-mass blue stars (e.g. Wilson, Scoville, & Rice 1991). In this paper we present $UBV$ photometry of the OB associations which power the two brightest giant HII regions in M33 with the aim of determining the star formation history and efficiency in these two unusual regions. The data reduction and identification of the OB associations are discussed in §2. The ages and star formation histories of the OB associations are derived in §3 and the implications for the structure of the interstellar medium are discussed. The star formation efficiencies are derived in §4. The paper is summarized in §5.

## 2. Observations and OB Association Analysis

$UBV$ observations of NGC 604 and NGC 595 were obtained during 1988 October 9-13 at the Palomar 60 in. telescope. The seeing was $1.6''$ for all images except the $B$ image of NGC 595, for which the seeing was $1.2''$ (chip scale $0.235''$ pixel$^{-1}$), and the nights were not photometric. The data were reduced using the ALLFRAME routine in the data reduction program DAOPHOT (Stetson 1987). Comparison of the ALLFRAME identifications with the Hubble Space Telescope images of Drissen et al. (1993) shows that ALLFRAME does a remarkable job of identifying stars correctly despite the relatively poor seeing and crowded fields. Although a few close neighbors cannot be separated using our photometry, the majority of the stellar identifications match stars seen in the HST images.

The data were calibrated by comparison with the data of Wilson (1991, hereafter W91), which were obtained during the same observing run. For NGC 595, the CCD frame overlapped directly with the W91 data set, and so could be easily calibrated. The total systematic uncertainties in the NGC 595 data are 0.05 mag in $V$, 0.05 mag in $B$, and 0.12 mag in $U$; most of the uncertainty in the $U$ magnitudes are due to uncertainty in the intrinsic calibration equation (W91). For NGC 604, four additional overlapping CCD frames were reduced and used to provide a transfer from the W91 data to the NGC 604 field. The total systematic uncertainties in the data are thus somewhat larger (0.07 mag in $V$, 0.09 mag in $B$, and 0.15 mag in $U$). The photometry for the individual stars in NGC 595 and NGC 604 is given in Tables 1 and 2.

OB associations in each of the two fields were identified using a "friends of friends" algorithm (W91), which places all stars that lie within one "search radius" of each other in the same association. We used the total number of stars bluer than $B - V = 0.4$ mag and brighter than $V = 21$ mag to calculate the search radius used to identify the OB associations (29 pixels or 26 pc for both regions). As expected, rich OB associations were identified at the location of each HII region. NGC 595 and NGC 604 each contain $\sim 50$ bright blue stars. Each OB association has a radius of $\sim 40$ pc, and thus the O star surface density is significantly higher than the average surface density in OB associations in the inner kiloparsec of M33 (W91). Figure 1 shows the OB association members and boundaries identified for each HII region.

The color-magnitude diagrams for the two OB associations are shown in Figure 2. The average reddening towards each association was determined by comparing the average color of the main sequence to the canonical color of unreddened O stars ($B - V = -0.3$).

The average reddening for NGC 595 is $E(B - V) = 0.25 \pm 0.06$ mag and for NGC 604 $E(B - V) = 0.35 \pm 0.10$ mag, where the uncertainties are simply the systematic uncertainties from the calibration. The much broader main sequence in NGC 604 suggests that there is significantly more differential reddening in NGC 604 than in NGC 595, which is consistent with the higher molecular gas content of NGC 604 (Wilson & Scoville 1992).

## 3. Ages and Star-Forming Histories of the HII Regions

The color-magnitude diagrams for the two associations are dominated by blue stars extending as bright at $V = 16 - 17$ mag. We can estimate an age for the most recent episode of star formation by determining the most massive star that is still on the main sequence. In NGC 595, there is an obvious break in the color-magnitude diagram at $V \sim 19$ mag. Of the 9 stars brighter than $V = 19$ mag, 5 are Wolf-Rayet stars identified by Drissen et al. (1993), while one likely includes the flux from two close neighbors that are resolved in the HST image. Two of the remaining stars have $V \sim 17.1$ mag, $B - V = 0.04 - 0.12$ mag, and values for the reddening-free parameter $Q = (U - B) - 0.72(B - V)$ of $-0.55$ to $-0.7$ mag. Their location relative to the evolutionary tracks from Maeder & Meynet (1988) suggests that these stars are most likely core-helium burning supergiants of initial masses $\sim 25$ M$_\odot$. The evolutionary status of the final star ($V \sim 19.2$ mag, $B - V \sim 0$ mag) is unclear: it could be a 40 M$_\odot$ star on the main sequence or a line-of-sight or intrinsic binary star. Since most of the stars brighter than $V = 19$ mag appear to be evolved stars, we estimate the turn-off to lie at $V = 19$ mag or $M_v = -6.3$ mag at the distance of M33 (0.79 Mpc, van den Bergh 1991). The least massive star which reaches $M_v = -6.3$ mag before leaving the main sequence has a mass of $\sim 25$ M$_\odot$ and a main sequence lifetime of $\sim 7$ Myr. Thus the age of the most recent episode of star formation in the OB association in NGC 595 is $\sim 7$ Myr.

The color-magnitude diagram of NGC 604 shows a much less clear break than that of NGC 595, perhaps due to the four times greater crowding in this OB association, although there does seem to be an increase in the number of blue stars fainter than $V \sim 18$ mag. Five of the six blue stars ($-0.1 < B - V < 0.2$ mag) with $V < 17$ mag are Wolf-Rayet or variable stars (Drissen et al. 1993). Of the 13 remaining stars brighter than $V = 17.7$ mag, 1 is a Wolf-Rayet star and 4 are clearly double in the HST images. There are again 2 very bright stars with $V = 16.1 - 16.7$ mag and $B - V = 0.2$ mag that are consistent with being evolved core-helium burning supergiants of initial masses $\sim 40$ M$_\odot$. The remaining 6 stars lie within 0.7 mag of the "turn-off" and could be either line-of-sight or intrinsic binaries. Thus we feel that the turn-off lies at $V \sim 18$ mag or $M_v = -7.6$ mag, which corresponds to a mass limit of $\sim 40 - 60$ M$_\odot$. Thus the age of the most recent episode of star formation in the OB association in NGC 604 is $\sim 4 - 5$ Myr, or 2-3 Myr younger than that of NGC 595.

It is clear from an inspection of Figure 2 that the location of the main sequence turnoff is less clearly defined in NGC 604 than in NGC 595. Thus it is possible that the turnoff in NGC 604 lies even fainter than $V \sim 18$ mag. However, there are several pieces of evidence which suggest that NGC 604 must be significantly younger than NGC 595. The brightest stars in NGC 604 are roughly a magnitude brighter than those in NGC 595. In addition, the brightest Wolf-Rayet stars in NGC 604 are also roughly a magnitude brighter than those in NGC 595 (Drissen et al. 1993). Finally, if the main sequence turnoff in NGC 604 were to lie as faint as that in NGC 595 ($V = 19$ mag), there are roughly 20 stars in NGC 604 which would lie more than 0.7 mag above the turnoff and hence would need to be unresolved triple or quadruple systems. Thus, although the precise location of the turnoff and hence the age of NGC 604 is more uncertain than for NGC 595, the evidence suggests that the OB association in NGC 604 is somewhat younger than that in NGC 595.

The presence of evolved red supergiants with progenitor masses below the main sequence turnoff is evidence for a prior episode of star formation. Inspection of Figure 2 shows that NGC 604 contains two faint red stars. However, these stars lie deep in the crowded heart of the HII region and we believe the photometry is not reliable for these stars; for example, their $Q$ values are unphysically blue. We would argue, therefore, that NGC 604 has had only one episode of massive star formation in the last 10-15 Myr. On the other hand, NGC 595 contains 3 red stars that are likely to have progenitor masses $< 15$ M$_\odot$, and thus ages $> 12$ Myr. Can these stars be extremely reddened O stars? True red supergiants have $U - B \sim 1.5$ mag and thus we would not expect to measure $U$ fluxes for these stars. Two of the stars have no $U$ magnitudes, while the third has $U = 21.4 \pm 0.2$ mag. As there is no obvious stellar peak at the position of this third star in $U$ while the background has a strong gradient due to the nearby HII region, we believe this $U$ measurement to be an upper limit. Since these stars are $\sim 1$ mag fainter than the red supergiants identified in OB associations in M33 by Regan & Wilson (1993), it is barely possible that they are in fact heavily reddened ($E(B - V) = 2 - 2.5$ mag) O stars. However, we feel that they are more likely to be evolved low-mass stars. The presence of these stars indicates that NGC 595 has undergone at least two episodes of star formation.

Are the numbers of evolved stars consistent with the ages we have estimated? To estimate the number of evolved stars, we assume a mass function $N(m) \propto m^{-2.5}$ for stars above 10 M$_\odot$ (the modified Miller-Scalo initial mass function, Kennicutt 1983). If we correct the counts for the NGC 595 OB association for incompleteness from $V = 19.5 - 21$ mag by the factors given in W91, we find a completeness-corrected population of 79 stars brighter than $V = 21$ mag. For an age of 7 Myr, $V = 21$ mag corresponds to a mass of $\sim 15$ M$_\odot$. Assuming that the evolved stellar lifetime is approximately 10% of the main sequence lifetime, we should only see evolved stars with main-sequence ages of 6.3 Myr or $\sim 30$ M$_\odot$ (interpolating ages from W91). Integrating the mass function between 15 and 25 M$_\odot$ and 25 and 30 M$_\odot$ shows that we would expect roughly 18% of the stars in NGC 595 to be evolved stars. Since there are 11 Wolf-Rayet stars identified in this region and we have identified 2 additional possible evolved stars, we observe an evolved star fraction of 16%, consistent with the assumed mass function and age.

For NGC 604, the counts are obviously very incomplete below 19.5 mag, and so we will normalize by the 23 blue stars brighter than $V = 19$ mag that are not Wolf-Rayet stars or candidate core-helium stars. For an age of 4.8 Myr, $V = 19$ mag corresponds to $\sim 35$ M$_\odot$. We would expect to see evolved stars that had main-sequence ages of 4.3 Myr or $\sim 50$ M$_\odot$. Integrating the mass function gives an expected evolved star fraction of $\sim 52\%$. Alternatively, if NGC 604 is as young as 3.6 Myr, $V = 19$ mag corresponds to a mass of 40 M$_\odot$, and we would expect to see evolved stars with main-sequence ages of 3.2 Myr or up to 90 M$_\odot$. Integrating the mass function for this age gives an expected evolved star fraction of 36%. The values for either age are in good agreement with the 14 Wolf-Rayet stars (plus perhaps 2 core-helium stars identified in this study) known in NGC 604 (Drissen et al. 1993).

What these numbers do not show clearly is the larger Wolf-Rayet to O star ratio in NGC 595. If we compare the observed number of Wolf-Rayet stars to the number of O stars with masses between 15 and 25 M$_\odot$ in NGC 595 (the observed numbers of blue stars on the main sequence corrected for incompleteness), we find a Wolf-Rayet to O star ratio of 0.17 in NGC 595. With our assumed mass function and the observed number of blue stars, integrating from 15 M$_\odot$ to the main sequence turnoff of 40-60 M$_\odot$ in NGC 604 gives a Wolf-Rayet to O star ratio of 0.04-0.07. For comparison, Drissen et al. (1993) measured ratios of 0.3 and 0.1, respectively. They suggested that NGC 595 was created in a single burst, while NGC 604 had undergone more than one episode of star formation, although they did note that differences in the ages of the bursts could cause the different WR/O ratios. Our age data clearly show that it is NGC 595 that has undergone multiple episodes of star formation. Hence the larger WR/O ratio in NGC 595 is a simple consequence of its older age: there are fewer O stars, because more have evolved off the main sequence, and there are relatively more Wolf-Rayet stars, because they are coming from more common lower mass progenitors. Indeed, the turn-off mass of NGC 595 is close to the lower mass limit for forming a Wolf-Rayet star according to recent models (Maeder et al. 1994).

The different star-formation histories of these two regions provide an explanation for their different gas environments (Wilson & Scoville 1992). NGC 595 contains comparable masses of atomic and molecular hydrogen, while molecular gas dominates the interstellar medium in NGC 604. The longer star-forming history of NGC 595 would allow it to photo-dissociate a larger fraction of its initial molecular gas reservoir. Indeed, although NGC 595 contains two molecular clouds (Wilson & Scoville 1992), they are relatively small compared to those in NGC 604 and to molecular clouds in the inner disk of M33. Thus NGC 595 is likely to be in the process of blowing itself apart, while NGC 604 has enough of a molecular gas reservoir that it may undergo another epoch of star formation. The data for NGC 595 show that a *large* episode of massive star formation is sufficient to destroy giant molecular clouds on timescales $> 2 \times 10^7$ yr. However, the O star surface density is higher than average for OB associations in M33 and the total number of O stars involved is also unusually high. Thus it is not clear whether *normal* episodes of star formation are sufficient to destroy molecular clouds or whether molecular clouds can survive many episodes of star formation. The fact that the region around NGC 595 has undergone two episodes of star formation in the last 15 Myr, as well as the presence of molecular clouds very close to NGC 604, shows that some molecular clouds in a 100 pc diameter region can survive at least one intense episode of star formation.

## 4. Star Formation Efficiency

We can obtain a good estimate of the star formation efficiency (defined to be mass of stars formed per mass of gas) in these giant HII regions by counting the number of massive stars that are present. We again adopt the modified Miller-Scalo initial mass function $(N(m) \propto m^{-2.5}$ for 1 $M_\odot < m < 100$ $M_\odot$, $N(m) \propto m^{-1.4}$ for 0.1 $M_\odot < m < 1$ $M_\odot$). For NGC 595, which contains 66 stars between 15 and 25 $M_\odot$, the total mass of stars between 10 and 100 $M_\odot$ formed in the most recent burst is 4600 $M_\odot$. The current molecular gas mass is $\sim 4 \times 10^5$ $M_\odot$, with $5.5 \times 10^5$ $M_\odot$ in the form of atomic hydrogen (Wilson & Scoville 1992; gas masses corrected for calibration error as described in Thornley & Wilson 1994). If we assume that all the gas was molecular at the time of the initial burst of star formation and include a factor of 1.36 to account for elements heavier than hydrogen, then the star formation efficiency is 0.3% for stars greater than 10 $M_\odot$, and 2.4% over the whole initial mass function. If we assume that the previous burst of star formation formed a similar mass of stars, the total star formation efficiency over the lifetime of the HII region would be $\sim$5%. NGC 595 has relatively little molecular gas left, which makes it unlikely that it could power a third burst of star formation of similar size. Thus this efficiency of 5% is likely to be the total star formation efficiency achieved by this complex in its lifetime. For NGC 604, , the total mass of stars formed in the initial burst between 10 and 100 $M_\odot$ is 17,000 $M_\odot$ if it is 4.8 Myr old and 8300 $M_\odot$ if it is 3.6 Myr old. The masses of molecular and atomic gas observed are $2.5 \times 10^6$ $M_\odot$ and $1.6 \times 10^6$ $M_\odot$, which gives a star formation efficiency of 0.14-0.3% for massive stars and 1.0-2.1% for stars of all masses. Since NGC 604 still contains a considerable reservoir of molecular gas, it seems likely that it will undergo at least one more burst of star formation, which would result in a total star formation efficiency comparable to that of NGC 595.

How do these star formation efficiencies compare to those of molecular clouds in the Milky Way and the average value in the disk of M33? Star formation efficiencies for nearby Galactic molecular clouds range from $\sim$ 1% for the Taurus-Auriga, Ophiuchus, and Orion A clouds to 3-4% for the Orion B molecular cloud (Evans & Lada 1991). These star formation efficiencies are measured over the current lifetime of the cloud, and thus the final star formation efficiencies may be somewhat larger. Thus the star formation efficiencies in the two giant HII regions in M33 are comparable to those in the Orion B molecular cloud. The high-mass star formation rate in the inner disk of M33 has been estimated to be $\geq 0.0042 \pm 0.0018$ M$_\odot$ yr$^{-1}$ from counts of the blue stars (Wilson, Scoville, & Rice 1991). Thus the total mass of stars formed with masses > 10 M$_\odot$ over the last 25 Myr (the main sequence lifetime of a 10 M$_\odot$ star) is $10^5$ M$_\odot$. The total gas reservoir (molecular + atomic) in the same region is $5 \times 10^7$ M$_\odot$, which gives a high-mass star formation efficiency of 0.2% and a total star formation efficiency of 1.4%. Thus the total star formation efficiency in NGC 595 is 3 times larger than the average in the disk of M33, while the efficiency in NGC 604 is currently comparable or just slightly higher.

Star formation "efficiencies" in the form of gas depletion times (the mass of gas divided by the current star formation rate) have been calculated previously for the disk of M33 (Wilson et al. 1991) and for the two giant HII regions (Wilson & Scoville 1992) by comparing the star formation rate obtained from the H$\alpha$ luminosity (Kennicutt 1983) with the amount of gas present. This comparison suggested that the giant HII regions had a star formation efficiency that was an order of magnitude larger than that of the inner disk, a result which disagrees with the current results. However, since our current understanding of the OB associations associated with these HII regions suggests that the stars originated in one or two bursts of star formation, we cannot simply use the formula given by Kennicutt, which is appropriate only for a region that is forming stars at a constant rate over a time at least as long as the lifetime of a 10 M$_\odot$ star ($\sim 25$ Myr, Schaller et al. 1992). For two regions with the same mass of gas and the same star formation efficiency, the region which formed all its stars in a single very young burst will produce a larger Lyman continuum luminosity than the region which formed its stars at a constant rate over 10-20 Myr. The steady-state region will have relatively fewer of the most massive stars, because of their short main-sequence lifetimes, and since the Lyman continuum luminosity scales very non-linearly with the stellar mass, these are the stars that produce most of the ionizing photons. As the burst ages, its Lyman continuum luminosity will decrease as the most massive stars evolve off the main sequence, and there will be some age at which the luminosity of the burst will fall below that of the steady-state region.

We have estimated the Lyman continuum flux that would be produced by a single burst of star formation of various ages and by a steady-state population. We use the models of Maeder & Meynet (1988) to obtain effective temperature on the zero-age main sequence and main sequence lifetime as a function of mass, the models of Panagia (1973) for the Lyman continuum flux per star as a function of effective temperature on the zero-age main sequence, and adopt the modified Miller-Scalo initial mass function. This set of assumptions produces a steady-state high-mass star formation rate that is a factor of 3 smaller than that of Kennicutt for the same Hα luminosity, presumably because the Lyman continuum flux over the star's lifetime is overestimated by adopting the zero-age main-sequence luminosity. The Lyman continuum luminosities in the burst models are 3 times larger at 3.6 Myr, 50% larger at 4.8 Myr, and 60% smaller at 7.1 Myr than the steady-state model. Although the exact figures are not highly accurate for the reasons given above, they do illustrate the increased luminosity of a young burst of star formation and the decrease in the luminosity of the burst with time. The exact age at which a burst will fade below the steady-state model will depend on the value of the upper mass cutoff to the initial mass function, since the most massive stars contribute a disproportionately large fraction of the ionizing photons. The dependence of the zero-age Lyman continuum luminosity on the slope and upper mass cutoff of the initial mass function is illustrated in Kennicutt & Chu (1988).

We can now attempt to determine why the Hα luminosities of these giant HII regions are so large compared to normal Galactic molecular clouds or average regions in the disk of M33. These two giant HII regions have Hα luminosity to gas mass ratios that are an order of magnitude larger than those of the inner disk of M33 (Wilson & Scoville 1992). This difference can most likely be explained by two factors: the somewhat higher star formation efficiencies in the giant HII regions (up to a factor of 3 in NGC 595), and the increased Lyman continuum luminosity produced in a young co-eval burst region compared to a steady-state region with a similar gas mass and star formation efficiency. In contrast, in comparing these giant HII regions with Galactic molecular clouds the most obvious difference is the size of the gas reservoir involved: the Orion molecular clouds have molecular gas masses of $\sim 10^5$ $M_\odot$, while the giant HII regions have gas reservoirs of $1 - 4 \times 10^6$ $M_\odot$. This difference in the size of the gas reservoir is likely one reason why these giant HII regions appear so luminous: for a larger gas mass, the same star formation efficiency produces many more O stars. However, since the H$\alpha$ luminosity of Orion is 100-400 times less than that of NGC 604 or NGC 595 (Kennicutt 1984), the increased total mass is unlikely to be the only explanation for the high H$\alpha$ luminosities of the M33 HII regions. A slightly higher star formation efficiency or a younger and/or co-eval burst of star formation may also provide part of the explanation. Small number statistics may also play a role: with a total mass only one-tenth that of NGC 595, Orion would be able to form at most a handful of O stars, and hence would be less likely to form any of the most massive O stars which contribute a large fraction of the total ionizing radiation.

## 5.    Conclusions

We have used UBV photometry to identify the OB associations which power the two most luminous giant HII regions in M33, NGC 604 and NGC 595, to investigate the star formation histories and star formation efficiencies of these unusual regions. Color-magnitude diagrams of the two OB associations reveal a significant difference in the ages of the most recent star formation episode in these two regions. Based on the location of the turnoff, the OB association that powers NGC 595 has an age of 7 Myr, while the OB association that powers NGC 604 has an age of only 4-5 Myr. In addition, NGC 595 contains a handful of evolved, low-mass red supergiants, which is evidence for a prior episode of star formation $\sim 10 - 15$ Myr ago in this HII region. The number of Wolf-Rayet and other evolved stars present in each HII region is consistent with the estimated turn-off ages and the number of stars still on the main sequence. We confirm that NGC 595 has a Wolf-Rayet to O star ratio that is a factor of two larger than in NGC 604; we attribute this difference primarily to the larger age of NGC 595.

The different star formation histories of these two regions can account for their different gas environments. Since NGC 595 is older and has had a prior episode of star formation, it has been able to destroy most of its molecular gas reservoir through photo-dissociation, so that it now has more atomic than molecular gas. In contrast, the interstellar medium of NGC 604 is still dominated by molecular gas, which is more normal for star-forming regions. The data for NGC 595 suggest that a large, compact episode of massive star formation can destroy molecular clouds on timescales of $2 \times 10^7$ yr. However, since NGC 604 still contains molecular clouds, and NGC 595 has had at least two episodes of star formation, molecular clouds appear to be able to survive at least one intense episode of massive star formation. It is also not clear whether normal star formation episodes are sufficient to disrupt molecular clouds.

The star formation efficiencies (mass of stars per mass of gas) of these two HII regions are up to a factor of 3 larger than the average efficiencies in the inner disk of M33 or in Galactic molecular clouds, but are still only 2-5%. The very large H$\alpha$ luminosities of these regions appear to be due to this slightly increased star formation efficiency, the large mass of gas involved in the burst of star formation, and the tendency for a young co-eval stellar population to produce more ionizing photons than a steady state population of similar mass.

C. D. W. was partially supported by NSERC Canada through a Women's Faculty

Award and Research Grant.

Table 1: OB Association Members for NGC 595 (plano table, separate LaTeX file)

Table 2: OB Association Members for NGC 604 (plano table, separate LaTeX file)

---

This manuscript was prepared with the AAS LATEX macros v3.0.

TABLE 1

OB Association Members for NGC 595 (to $V = 20$ mag)

| Star | X | Y | V | $B-V$ | $U-B$ | $\sigma_V$ | $\sigma_B$ | $\sigma_U$ | notes[a] |
|------|-----|-----|-------|-------|-------|-------|-------|-------|----------|
| 1 | 154 | 618 | 16.74 | 0.40 | -0.42 | 0.01 | 0.01 | 0.03 | incl. WR11 |
| 2 | 136 | 645 | 17.04 | 0.04 | -0.67 | 0.01 | 0.01 | 0.01 | core He? |
| 3 | 150 | 602 | 17.09 | 0.12 | -0.47 | 0.01 | 0.01 | 0.02 | core He? |
| 4 | 147 | 616 | 17.36 | 0.10 | -0.74 | 0.02 | 0.02 | 0.02 | WR2A.2B |
| 5 | 155 | 601 | 17.91 | -0.05 | -1.01 | 0.03 | 0.03 | 0.01 | 2 stars |
| 6 | 118 | 573 | 18.16 | -0.03 | -0.50 | 0.01 | 0.01 | 0.02 | |
| 7 | 137 | 573 | 18.17 | -0.16 | -0.75 | 0.01 | 0.01 | 0.02 | WR4 |
| 8 | 95 | 566 | 18.26 | -0.16 | -0.64 | 0.01 | 0.01 | 0.01 | WR5 |
| 9 | 131 | 644 | 18.28 | -0.18 | -0.31 | 0.01 | 0.01 | 0.02 | WR1 |
| 10 | 129 | 570 | 19.07 | -0.06 | -0.85 | 0.02 | 0.01 | 0.01 | |
| 11 | 166 | 618 | 19.13 | 0.36 | -0.99 | 0.03 | 0.03 | 0.03 | |
| 12 | 138 | 616 | 19.18 | -0.02 | -0.81 | 0.03 | 0.04 | 0.02 | |
| 13 | 165 | 611 | 19.21 | 0.05 | -0.80 | 0.03 | 0.03 | 0.02 | |
| 14 | 129 | 602 | 19.23 | -0.10 | -0.93 | 0.02 | 0.01 | 0.02 | |
| 15 | 164 | 599 | 19.26 | 0.17 | -0.84 | 0.04 | 0.04 | 0.02 | |
| 16 | 95 | 596 | 19.28 | 1.73 | 0.41 | 0.04 | 0.05 | 0.13 | |
| 17 | 163 | 646 | 19.46 | 0.35 | -0.38 | 0.02 | 0.01 | 0.02 | |
| 18 | 124 | 570 | 19.52 | -0.05 | -0.52 | 0.02 | 0.02 | 0.02 | |
| 19 | 174 | 604 | 19.60 | 0.10 | -0.88 | 0.03 | 0.03 | 0.02 | |
| 20 | 170 | 601 | 19.63 | 0.05 | -0.91 | 0.04 | 0.05 | 0.02 | |
| 21 | 165 | 603 | 19.65 | 0.08 | -0.52 | 0.04 | 0.05 | 0.03 | |
| 22 | 131 | 642 | 19.65 | 0.00 | -2.05 | 0.05 | 0.05 | 0.01 | |
| 23 | 91 | 579 | 19.68 | 2.14 | ... | 0.02 | 0.21 | ... | |
| 24 | 164 | 591 | 19.68 | -0.10 | -0.93 | 0.04 | 0.04 | 0.01 | |
| 25 | 94 | 576 | 19.71 | -0.13 | -0.86 | 0.03 | 0.02 | 0.02 | |
| 26 | 141 | 606 | 19.77 | -0.20 | -0.91 | 0.09 | 0.05 | 0.03 | |
| 27 | 88 | 632 | 19.80 | 0.00 | -0.71 | 0.03 | 0.02 | 0.02 | |
| 28 | 144 | 611 | 19.88 | -0.03 | -1.70 | 0.16 | 0.15 | 0.02 | |
| 29 | 91 | 639 | 19.92 | -0.12 | -0.94 | 0.03 | 0.02 | 0.02 | |
| 30 | 155 | 606 | 19.92 | 0.07 | 0.28 | 0.09 | 0.07 | 0.25 | |
| 31 | 168 | 594 | 20.00 | 0.07 | -0.74 | 0.06 | 0.06 | 0.04 | |

[a]Wolf-Rayet and multiple identifications from Drissen et al. (1983)

TABLE 2

OB Association Members for NGC 604 (to $V = 18.5$ mag)

| Star | X | Y | $V$ | $B-V$ | $U-B$ | $\sigma_V$ | $\sigma_B$ | $\sigma_U$ | notes[a] |
|------|-----|-----|-------|-------|-------|------|------|------|----------|
| 1 | 177 | 321 | 15.97 | 0.15 | -0.84 | 0.01 | 0.01 | 0.01 | WR4A,4B |
| 2 | 162 | 328 | 16.09 | 0.21 | -0.96 | 0.01 | 0.01 | 0.01 | core He? |
| 3 | 206 | 312 | 16.29 | 0.00 | -1.06 | 0.02 | 0.01 | 0.01 | WR2A,2B |
| 4 | 207 | 347 | 16.29 | 0.50 | -1.07 | 0.03 | 0.03 | 0.00 | multiple |
| 5 | 183 | 374 | 16.43 | 0.14 | -0.90 | 0.01 | 0.01 | 0.01 | WR6 |
| 6 | 173 | 326 | 16.49 | 0.45 | -1.20 | 0.03 | 0.06 | 0.01 | double |
| 7 | 186 | 322 | 16.69 | -0.05 | -0.93 | 0.03 | 0.02 | 0.01 | WR3 |
| 8 | 216 | 366 | 16.69 | 0.19 | -0.80 | 0.01 | 0.02 | 0.01 | core He? |
| 9 | 191 | 301 | 16.85 | -0.02 | -1.08 | 0.02 | 0.01 | 0.01 | V1 |
| 10 | 205 | 304 | 17.00 | 0.12 | -0.93 | 0.02 | 0.01 | 0.01 | WR1 |
| 11 | 178 | 312 | 17.05 | 0.22 | -0.79 | 0.03 | 0.02 | 0.02 | double |
| 12 | 211 | 308 | 17.15 | 0.42 | -1.18 | 0.02 | 0.03 | 0.01 | |
| 13 | 190 | 314 | 17.17 | 0.39 | -0.78 | 0.05 | 0.03 | 0.02 | |
| 14 | 172 | 331 | 17.22 | 0.02 | -0.89 | 0.03 | 0.03 | 0.01 | |
| 15 | 179 | 305 | 17.25 | 0.06 | -1.18 | 0.02 | 0.03 | 0.01 | double |
| 16 | 200 | 303 | 17.27 | 0.68 | -0.60 | 0.02 | 0.02 | 0.02 | |
| 17 | 196 | 309 | 17.51 | 0.14 | -1.22 | 0.04 | 0.03 | 0.01 | |
| 18 | 206 | 296 | 17.61 | 0.48 | -1.00 | 0.05 | 0.03 | 0.02 | |
| 19 | 222 | 363 | 17.74 | 0.15 | -0.86 | 0.04 | 0.03 | 0.01 | |
| 20 | 193 | 286 | 17.77 | 0.36 | -1.12 | 0.02 | 0.02 | 0.02 | |
| 21 | 168 | 321 | 17.78 | 0.11 | -0.51 | 0.05 | 0.05 | 0.03 | |
| 22 | 210 | 340 | 17.80 | 0.25 | -0.64 | 0.03 | 0.04 | 0.01 | |
| 23 | 168 | 373 | 17.89 | 0.47 | -1.48 | 0.03 | 0.07 | 0.02 | |
| 24 | 217 | 357 | 17.90 | 0.65 | -1.32 | 0.03 | 0.10 | 0.02 | |
| 25 | 203 | 358 | 17.93 | 0.09 | -1.17 | 0.03 | 0.02 | 0.01 | |
| 26 | 226 | 382 | 17.96 | 0.71 | -0.35 | 0.02 | 0.02 | 0.03 | |
| 27 | 198 | 295 | 17.99 | 0.44 | -1.49 | 0.04 | 0.06 | 0.01 | |
| 28 | 223 | 354 | 18.00 | 0.07 | -0.77 | 0.02 | 0.02 | 0.02 | |
| 29 | 229 | 358 | 18.01 | 0.04 | -1.10 | 0.03 | 0.03 | 0.01 | |
| 30 | 184 | 312 | 18.06 | -0.46 | -0.94 | 0.06 | 0.02 | 0.02 | |
| 31 | 193 | 292 | 18.06 | 0.26 | -1.12 | 0.04 | 0.03 | 0.02 | |
| 32 | 183 | 344 | 18.14 | -0.02 | -0.99 | 0.02 | 0.02 | 0.02 | |
| 33 | 233 | 366 | 18.21 | 0.20 | -1.13 | 0.02 | 0.02 | 0.02 | |
| 34 | 210 | 318 | 18.24 | 1.41 | -2.48 | 0.04 | 0.29 | 0.02 | |
| 35 | 221 | 322 | 18.28 | 0.00 | -1.17 | 0.06 | 0.05 | 0.02 | |
| 36 | 183 | 365 | 18.33 | 0.54 | -0.62 | 0.09 | 0.04 | 0.02 | |
| 37 | 179 | 293 | 18.33 | 0.09 | -1.18 | 0.06 | 0.04 | 0.02 | |
| 38 | 226 | 367 | 18.34 | -0.01 | -0.97 | 0.03 | 0.02 | 0.03 | |
| 39 | 215 | 314 | 18.39 | -0.38 | -0.90 | 0.08 | 0.03 | 0.01 | |
| 40 | 187 | 286 | 18.41 | 0.02 | -1.09 | 0.04 | 0.05 | 0.01 | |
| 41 | 227 | 347 | 18.42 | 0.03 | -0.73 | 0.03 | 0.02 | 0.02 | |
| 42 | 242 | 362 | 18.44 | 0.26 | -0.91 | 0.04 | 0.04 | 0.02 | |
| 43 | 197 | 343 | 18.48 | 0.17 | -1.06 | 0.04 | 0.05 | 0.02 | |

[a]Wolf-Rayet and multiple identifications from Drissen et al. (1983)

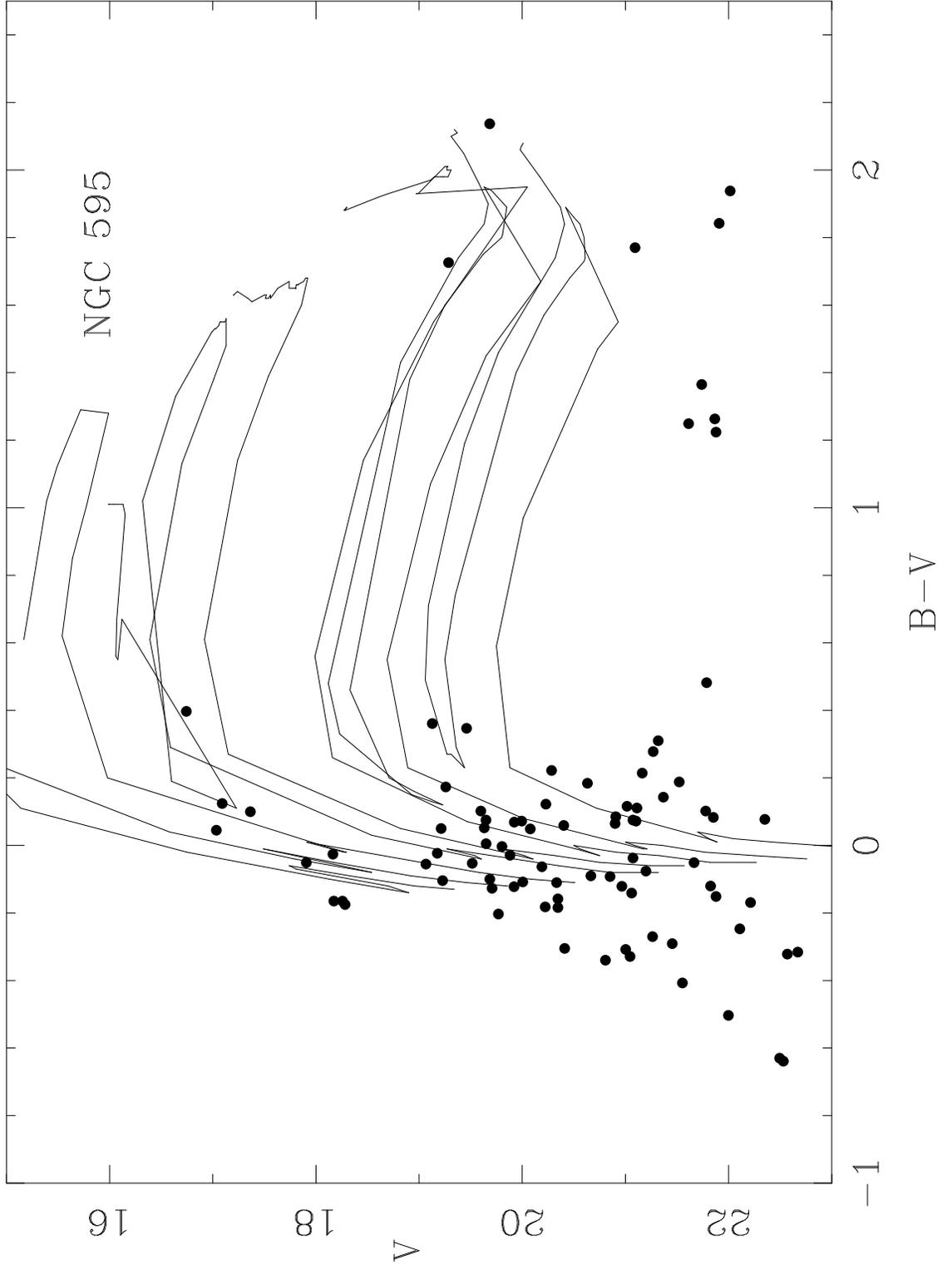

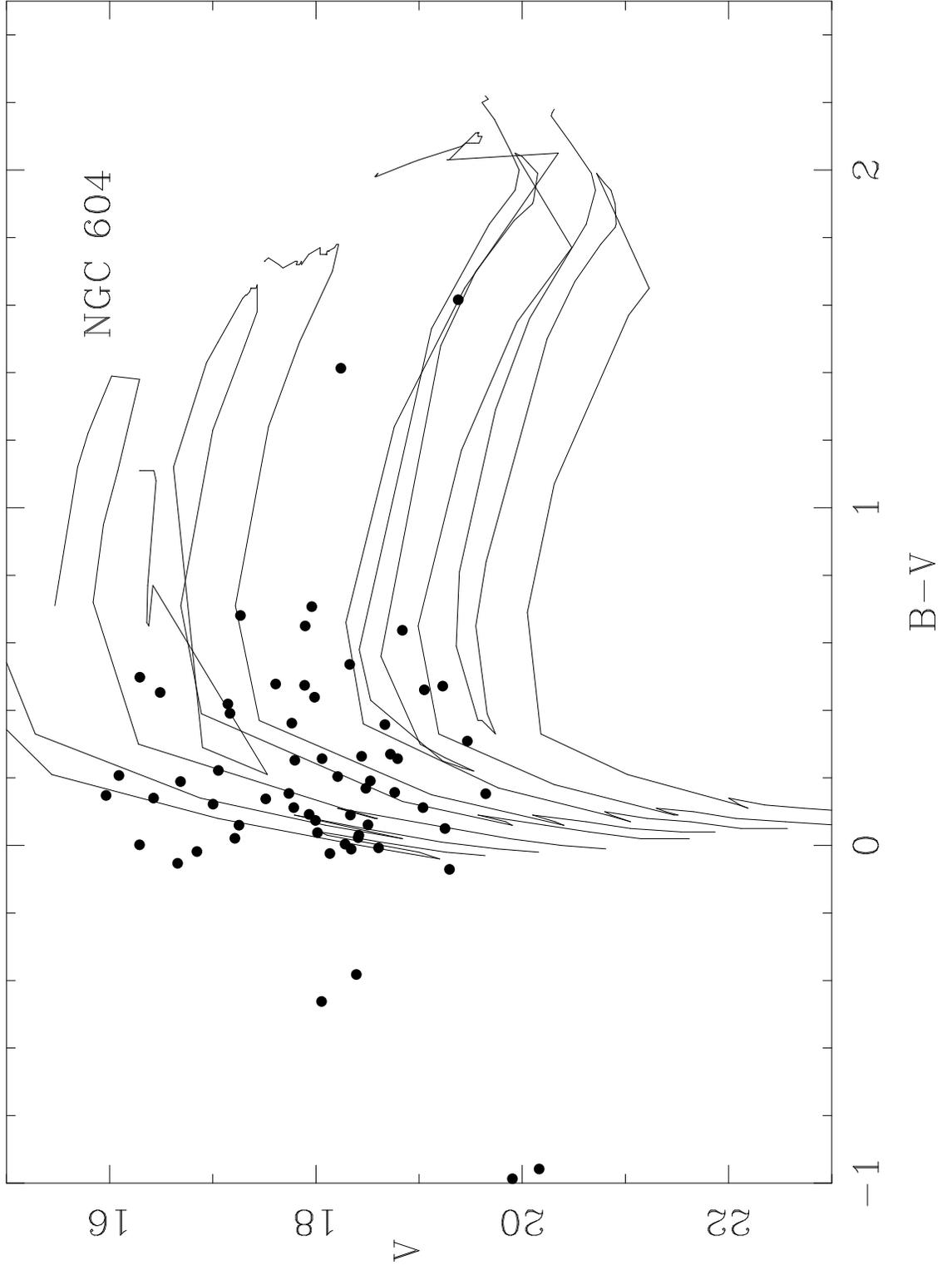

Fig. 1.— $V$ images of the OB associations in NGC 604 and NGC 595. All stars with $B - V < 0.4$ and $V < 21$ are marked with white squares. The boundaries of the associations are indicated by the heavy lines. North is at the top and east is to the left. For the stellar coordinates given in the Tables, X increases down the page, Y increases to the left, and the images are 400 by 400 pixels in size. (a) NGC 595. The upper left corner is (1,800). (b) NGC 604. The upper left corner is (1, 500). (Note: not available in electronic form.)

Fig. 2.— $V$ versus $B - V$ color-magnitude diagram of NGC 595 and NGC 604. The evolutionary tracks for 9, 12, 15, 20, 25, 40, 60, and 85 $M_{\odot}$ stars from Maeder & Meynet (1988) are overlaid. The adopted reddenings are $E(B - V) = 0.25$ mag for NGC 595 and $E(B - V) = 0.35$ mag for NGC 604.